\documentclass[fleqn,10pt]{wlscirep}
\usepackage[utf8]{inputenc}
\usepackage[T1]{fontenc}

\title{Optical cycling of MgF molecules within the hyperfine states in X(N=1) state}

\author[1]{Kikyeong Kwon}
\author[1]{Seunghwan Roh}
\author[1]{Youngju Cho}
\author[1]{Yongwoong Lee}
\author[1,*]{Eunmi Chae}
\affil[1]{Korea University, Department of Physics, Seoul, 02841, The Republic of Korea}

\affil[*]{echae@korea.ac.kr}


\begin{abstract}
We investigated the optical cycling effect of the $\mathrm{X}^2\Sigma(v=0,\ N=1^-) - \mathrm{A}^2\Pi_{1/2}(v'=0,\ J'=1/2^+)$ band of MgF molecules, specifically the $\mathrm{P_1/Q_{12}(1)}$ transition, which serves as the main transition in the quasi-closed cycling scheme for the laser cooling.
A higher number of scattered photons was observed when all three frequency components of the $\mathrm{P_1/Q_{12}(1)}$ transition were simultaneously applied using acousto-optic modulators (AOMs). 
Optimal conditions were identified by scanning the detuning of frequency components, the laser beam power ratio, and the total laser beam power, and the results were confirmed through rate equation simulations.
Under these optimized conditions, and with an applied magnetic field, the scattering rate was enhanced by approximately a factor of six.
These results refine the implementation of optical cycling in MgF and lay the groundwork for laser slowing and magneto-optical trapping (MOT) experiments.

\end{abstract}
\begin{document}
\flushbottom
\maketitle
\thispagestyle{empty}

\section*{Introduction}

Over the past decade, ultracold molecules have become promising candidates for fundamental physics research, including parity-time violation (PTV)~\cite{Hudson2011, Baron2014, Cairncross2017, Andreev2018} and dark matter searches~\cite{Kozyryev2021}, as well as quantum chemistry~\cite{Ospelkaus2010, Gregory2019}, due to their rich internal energy structures and intrinsic electric dipole moments. 
Moreover, due to their long coherence time of rovibrational states and capability for long-range dipole-dipole interactions, these molecules have recently emerged as essential tools in quantum computation and simulation~\cite{Chae2024, Cornish2024QP, Bigagli2024_NaCsBEC}

To utilize these unique features, achieving ultracold temperatures is crucial.
Fortunately, various cooling techniques, including laser cooling, have already been extensively developed for neutral atoms, and these established methods can be applied to molecular systems.
Such laser cooling techniques are increasingly being applied to hydrogen-like diatomic molecules, particularly alkaline-earth monofluorides, which exhibit diagonal Franck-Condon factors (FCFs) and possess a single valence electron, enabling effective adaptation of atomic cooling methods.

In particular, extensive technical developments have been achieved with CaF molecules, including their recent loading into tweezer arrays, and the demonstration of spin-exchange interactions and entanglement~\cite{Bao2023_CaF_spin, Anderegg2018_CaF_ODT, Bao2024_CaF_RSC, Lu2024_CaF_RSC, Li2024_CaF_BlueMOT, Holland2023_CaF_Entanglement, Zhelyazkova2014_CaF_Cooling}. 
SrF molecules have also been successfully captured in optical dipole traps (ODTs) and cooled down to 40 $\mu \rm{K}$ using sub-Doppler cooling techniques~\cite{Jorapur2023_SrF_MOT, Truppe2017_SrF_subDcool}. 
Furthermore, various molecules, such as YbF, AlF, and BaF, are being actively studied using laser slowing and magneto-optical trapping (MOT) for precision measurements~\cite{Tarbutt2025_YbF_EDM, PadillaCastillo2025_AlF_MOT, Zeng2024_BaF_MOT, Zeng2025_BaF_BlueMOT}. 
Additionally, novel cooling methods uniquely suited to diatomic molecules, such as radio-frequency (RF) MOTs~\cite{Anderegg2017_CaF_RFMOT}, blue-detuned MOTs~\cite{Yu2024_CaF_RF_convMOT}, and microwave shielding~\cite{Karman2018_MW_Shield, Anderegg2021_MW_Shield}, are actively being developed.

MgF shares a similar energy structure and electric dipole moments with other alkaline-earth monofluorides but offers additional advantages for laser cooling due to its lighter mass and ultraviolet (UV) transition wavelengths.
Consequently, laser slowing and MOT for MgF molecules have been proposed~\cite{Yan2022_MgF_slowing_simulation, Rodriguez2023_MgF_MOT_simulation}. 
The hyperfine-resolved spectroscopy studies were performed~\cite{Doppelbauer2022_MgF_hyperfine_spectroscopy}, and measurements of branching fractions and decay rates of excited states involved in quasi-closed cycling transitions have also been reported~\cite{Norrgard2023_MgF_FCF}. 
Recently, laser-induced fluorescence (LIF) spectroscopy for cooling lines, such as $\mathrm{X}(v=1)-\mathrm{A}(v'=0)$, $\mathrm{X}(v=1)-\mathrm{B}(v'=0)$, and $\mathrm{X}(v=2)-\mathrm{A} (v'=1)$, was conducted~\cite{Pilgram2024_MgF_coolingline_spectrscopy}.

In preparation for future magneto-optical trapping (MOT) experiments with MgF, obtaining a sufficient number of scattered photons through closed-cycling transitions is essential. 
This study aims to optimize the photon scattering rate for the rotationally closed transition ($N=1^- \to J'=1/2^+$) within the main transition of the $\mathrm{X}(v=0)-\mathrm{A}(v'=0)$ band, which exhibits the strongest scattering.

In previous studies~\cite{Gu2022_MgF_OC, Pilgram2024_MgF_coolingline_spectrscopy}, electro-optic modulators (EOMs) were utilized to generate frequency components corresponding to transitions from each hyperfine state of the ground state, thereby achieving a rotationally closed transition. 
When only the main laser was applied without vibrational repump lasers, an enhancement of the LIF signal by approximately a factor of 2.6 was reported~\cite{Gu2022_MgF_OC}. 
However, since EOMs produce frequency components evenly spaced by the modulation frequency, and the power ratio between these components is symmetrically determined by the modulation depth, this approach significantly limits the accessible parameter space for the rotationally closed transition of the main transition.

In this study, we overcome these limitations by employing acousto-optic modulators (AOMs), which allow for independent control of the detuning and power of each frequency component. 
This enabled exploration of a broader parameter space. 
Additionally, by applying a magnetic field at a tilted angle relative to the laser polarization, we induced dark state mixing through Larmor precession, effectively mitigating these dark states.

\section*{Rotationally Closed Transition of $\mathbf{X(0)-A(0)}$ band}

\begin{figure}[ht]
    \centering
    \textbf{(a)}\hspace{0.45\linewidth} \textbf{(b)} \\ 
        \includegraphics[width=\linewidth]{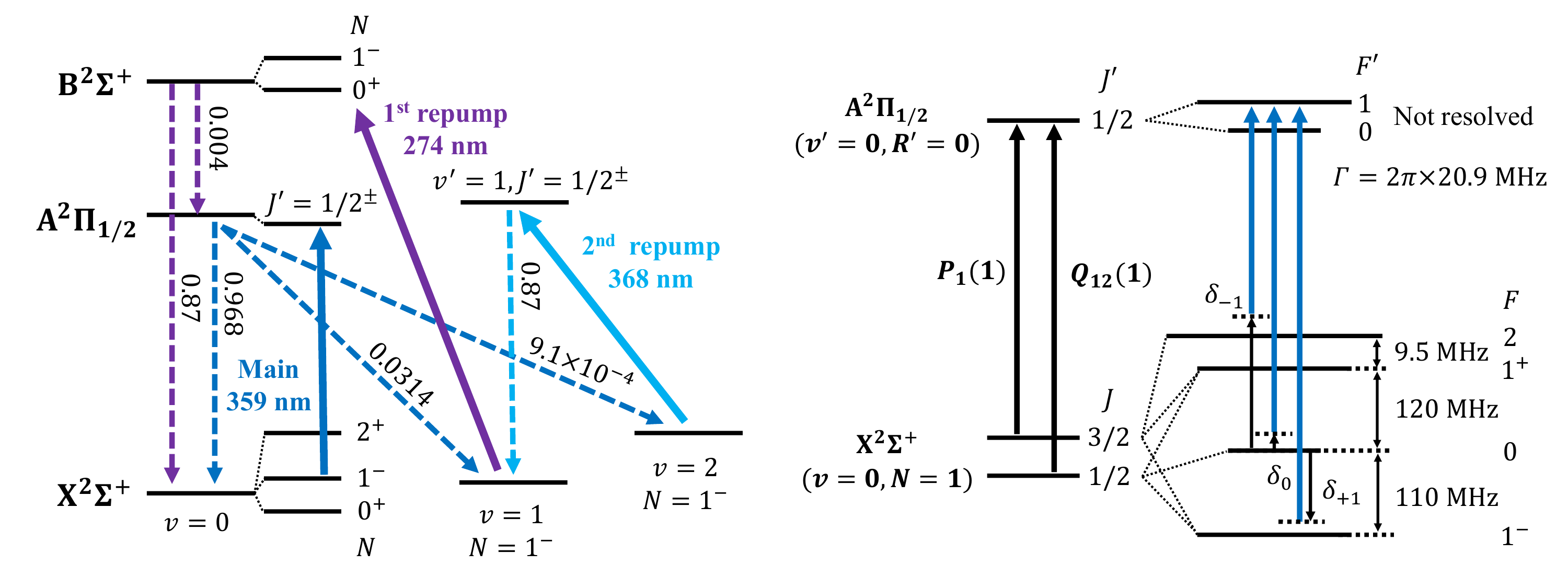}
    \caption{(a) Quasi-closed cycling transitions for the rovibrational branches of MgF. Solid lines represent the transitions and their wavelengths, while dashed lines indicate dominant vibrational decay channels with their branching fractions. 
    (b) Fine and hyperfine energy levels of the $\mathrm{X}(v=0,\ N=1^-)-\mathrm{A}(v'=0,\ J'=1/2^+)$ band. The black arrows indicate the rotationally closed transitions, $\mathrm{P_1/Q_{12}(1)}$, while the blue arrows represent the frequencies of the excitation lasers for $\mathrm{P_1/Q_{12}(1)}$ transition. The detuning of each laser is referenced to 834294485 $\mathrm{MHz}$, which is the frequency of the transition from $F=0$ state.}
    \label{fig:energydiagramtotal}
\end{figure}

The transitions forming a quasi-closed cycling transition must each satisfy rotationally closed conditions to achieve the required number of scattered photons for laser cooling.
Parity and angular momentum selection rules allow the transitions between states with different rotational quantum numbers.
For the $\mathrm{X}(v=0)-\mathrm{A}(v'=0)$ band, a rotationally closed transition can be realized by selecting the $N=1^- \to J'=1/2^+$ transition.(Figure.\ref{fig:energydiagramtotal}(a))

However, achieving a fully closed cycling transition is no longer feasible using a laser with a single frequency component due to the hyperfine structure of MgF. 
Consequently, all transitions from relevant hyperfine states must be simultaneously driven to sustain optical cycling (Figure \ref{fig:energydiagramtotal}(b)).
These $J=3/2, 1/2 \to J'=1/2^-$ transitions are collectively designated as the $\mathrm{P_1/Q_{12}(1)}$ transitions in Fig. \ref{fig:energydiagramtotal}(b). 
Here, unprimed quantum numbers denote ground states, while primed quantum numbers refer to excited states.
In this experiment, we investigated the optical cycling properties of these transitions to maximize the scattering rate.

To simultaneously drive the $\mathrm{P_1/Q_{12}(1)}$ transitions, a total of six frequency components may appear to be required. However, due to the large natural decay rate $\Gamma = 2\pi \times 20.9 \ \mathrm{MHz}$ of the A state~\cite{Norrgard2023_MgF_FCF}, the hyperfine splitting between $F' = 1, 0$ in the excited state and between $F = 2, 1^+$ states in the ground cannot be resolved. Consequently, all transitions within the $\mathrm{P_1/Q_{12}(1)}$ manifold can be effectively driven using three frequency components centered at the transition from $|J=1/2,\ F=0\rangle$, with detunings of $(-125,\ 0,\ +110) \ \mathrm{MHz}$.

\begin{figure}[ht]
  \centering
    \textbf{(a)}\hspace{0.45\linewidth} \textbf{(b)} \\ 
        \includegraphics[width=\linewidth]{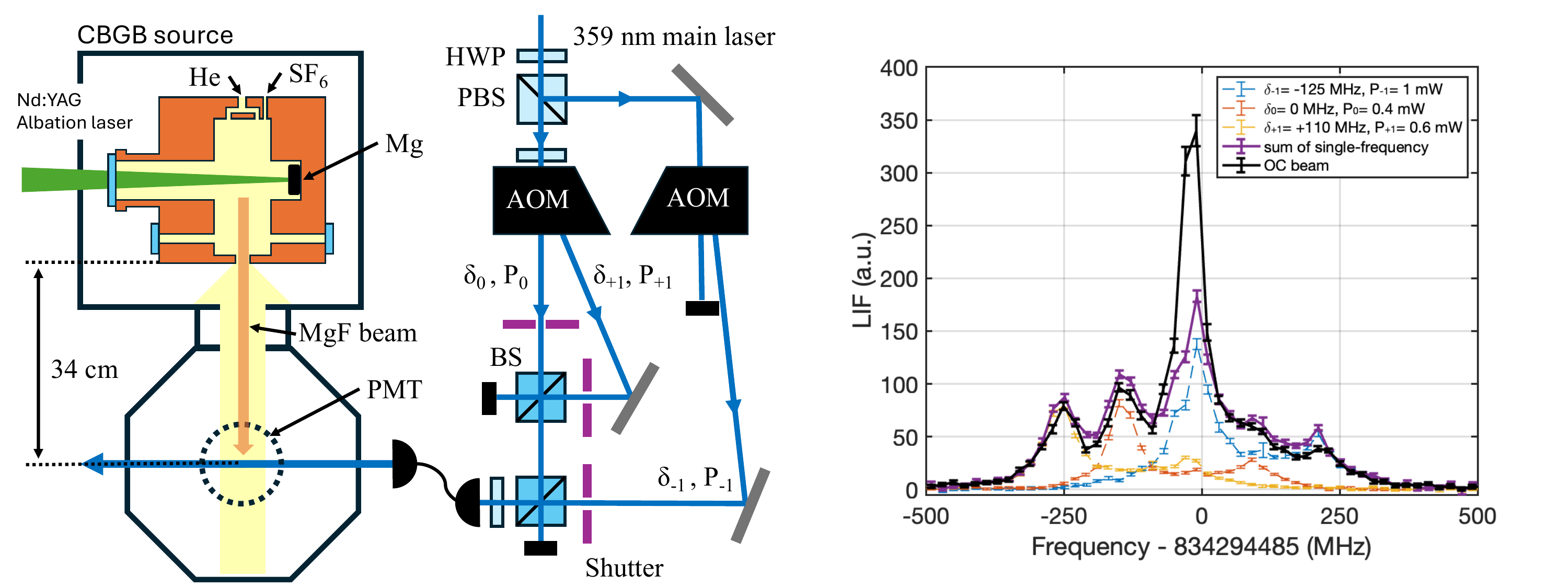}
    \caption{(a) Schematic diagram of the experimental setup. Two AOMs generate the frequency components of the OC beam, which is delivered to the chamber via an optical fiber. The molecular beam(yellow region), produced by the CBGB(Cryogenic buffer-gas beam) source, interacts perpendicularly with the laser beam approximately 34 ${\mathrm{cm}}$ downstream. The PMT records the LIF signal for 20 $\mathrm{ms}$ after ablation.
    (b) Spectra for the OC beam (solid black line) and single-frequency components (dashed line). LIF signals were obtained by scanning the main laser frequency while keeping $(\delta_i,\ P_i)$ of each frequency component fixed. The purple solid line represents the sum of single-beam spectra at each frequency. Near resonance, where all three frequency components address the hyperfine transitions, the optical cycling effect becomes evident.}
    \label{fig:Full spectrum and Exp set up}
\end{figure}

\section*{Experimental Setup}

As depicted in Figure~\ref{fig:Full spectrum and Exp set up}(a), MgF molecules are produced by ablating a Mg target with a 532 $\mathrm{nm}$ pulsed laser in a cryogenic system cooled below 4 $\mathrm{K}$. The ablated Mg atoms then react with $\mathrm{SF_6}$ gas to form MgF.
The molecules are then cooled to below 10 $\mathrm{K}$ through elastic collisions with He gas at 4 $\mathrm{K}$. 
After cooling, the molecular beam exits the cryogenic system as a collimated beam and propagates to the detection region located 34 $\mathrm{cm}$ downstream. The forward velocity of the molecular beam is centered at around 180 $\mathrm{m/s}$, with a width of approximately 100 $\mathrm{m/s}$.

Two AOMs with a resonant frequency range of 110 - 130 $\mathrm{MHz}$  were used to generate sidebands corresponding to the hyperfine structure of  $\mathrm{P_1/Q_{12}(1)}$ transitions.
By changing the amplitudes and frequencies of the RF source applied to the AOMs, the detunings ($\delta_{\pm1}$) and powers($ P_{0},\ P_{\pm1} $) of the carrier and sidebands were adjusted. 
The detunings $\delta_i$ are defined relative to the resonance frequency $f_0 = 834294485\ \mathrm{MHz}$, corresponding to the transition from the ground state $F=0$~\cite{Doppelbauer2022_MgF_hyperfine_spectroscopy}.
These frequency components were then combined using two beam splitters (BS), forming what we refer to as the optical cycling beam(OC beam), which is linearly polarized. 
The OC beam was then coupled into a polarization-maintaining fiber and delivered to the chamber, with a beam waist radius of 0.5 $\mathrm{mm}$.

A photomultiplier tube (PMT) detected the LIF signal when the molecular beam intersected the laser beam perpendicularly.
The time-trace signals were acquired for 20 $\mathrm{ms}$ after the ablation pulse to collect all the fluorescence from molecules with different velocities and departure times.

\section*{Result and Discussion}

Figure \ref{fig:Full spectrum and Exp set up}(b) presents the LIF spectra obtained by scanning the frequency of the main laser while keeping $\delta_i$ and $P_i$ fixed, clearly illustrating the effect of optical cycling induced by the three frequency components.
These spectra were acquired by selectively applying each frequency component individually by blocking other frequency components using mechanical shutters(single-frequency spectrum) or by applying the OC beam(OC spectrum).
In addition, the sum of individual single-frequency spectra was computed and used as a reference to evaluate the effect of optical cycling.
The single-frequency spectra (blue, orange, yellow dotted line in Fig. \ref{fig:Full spectrum and Exp set up}(b)), measured by applying each frequency component separately, exhibit three peaks corresponding to transitions from $F=2,1^+$, $F=0$, and $F=1^-$, from left to right.
Each of these spectra is shifted by $-\delta_i$ due to the modulation from the AOM, and the relative peak heights vary according to $P_i$.
The $\delta_i$ values were tuned to be resonant with the hyperfine structure, and the power ratio $P_i$ was set in proportion to the number of magnetic sublevels.

Compared to the single-frequency spectra, both the sum of single-frequency spectra(purple solid line in Fig. \ref{fig:Full spectrum and Exp set up}(b)) and the OC spectrum(black solid line in Fig. \ref{fig:Full spectrum and Exp set up}(b)) reveal two additional peaks arising from the presence of frequency sideband components.
The largest peak appears when the main laser is resonant with the transition from the $F=0$ state (i.e., at the zero point on the horizontal axis).
The increase in scattering rate due to the optical cycling effect can be inferred by comparing the peak heights of the sum of single-frequency spectra and the OC spectrum.
As shown in Figure \ref{fig:Full spectrum and Exp set up}(b), a higher scattering rate was observed when all three frequency components simultaneously addressed the corresponding transitions, forming a closed-cycling transition.
The following sections describe optimization of the laser parameters to maximize the scattering rate.
Under fully optimized conditions, the LIF signal obtained using the OC beam was up to three times greater than the sum of the single-frequency LIF signals in our experimental setup.

\subsection*{Detunings}
The detunings $\delta_{\pm1}$ were individually scanned by changing the modulation frequency of AOM, while $\delta_0$ was adjusted by shifting the frequency of the main laser.

Intuitively, each frequency component should be tuned to resonate with the transitions corresponding to the $\mathrm{P_1/Q_{12}(1)}$ transitions. 
However, in our setup, a single frequency drives both transitions from $F=2,1^+$, whose energy splitting is approximately 0.4 $\Gamma$.
As a result, the optimal value of $\delta_{-1}$ for maximum photon scattering should be examined.

\begin{figure}[ht]
    \centering
    \textbf{(a)}\hspace{0.28\linewidth} \textbf{(b)}\hspace{0.28\linewidth} \textbf{(c)} \\ 
        \includegraphics[width=\linewidth]{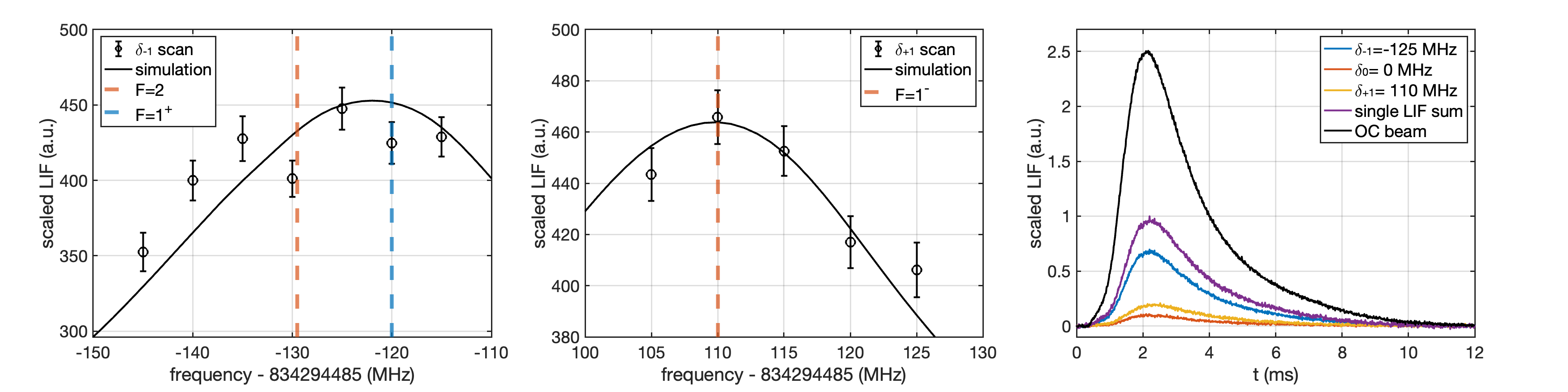}
    \caption{(a, b) The dependence of LIF from the OC beam as a function of (a)$\delta_{-1}$  and (b)$\delta_{+1}$, while keeping the other components fixed at (a) ($\delta_{0}=0$, $\delta_{+1}=110$) MHz and (b) ($\delta_{-1}=-125$, $\delta_{0}=0$) MHz, respectively. 
    Solid lines indicate simulation results, and dashed lines mark transition frequencies for $F=2$, $F=1^+$, and $F=1^-$. (c) Comparison of LIF time traces between the OC beam at the optimal detuning and the single-frequency beam. All experiments were conducted with a total laser beam power of 2 $\mathrm{mW}$, distributed in the ratio of $(P_{-1}:P_0:P_{+1}) = (5:2:3)$.}
    \label{fig:SingleFreqScan}
\end{figure}

The experimental results showed that the optimal detunings for $\delta_0$ and $\delta_{+1}$ were when they were resonant with the transitions from $F=0$ and $F=1^-$, respectively, as we expected. 
In contrast, the optimal detuning range for $\delta_{-1}$ was broader than that of $\delta_{0, +1}$, primarily due to the energy splitting between the $F=2$ and $F=1^+$ states. 
The most optimal detuning was found at $-125\ \mathrm{MHz}$, approximately at the midpoint of these two transitions.
Figure \ref{fig:SingleFreqScan}(c) shows the time trace signals at the optimal detunings.
The increase in scattering rate due to the optical cycling effect is evident, where the time trace of the OC beam LIF (black) exceeds that of the sum of single-frequency LIF signals (purple).

\subsection*{Power ratio}

\begin{figure}[ht]
    \centering
    \textbf{(a)}\hspace{0.45\linewidth} \textbf{(b)} \\ 
        \includegraphics[width=\linewidth]{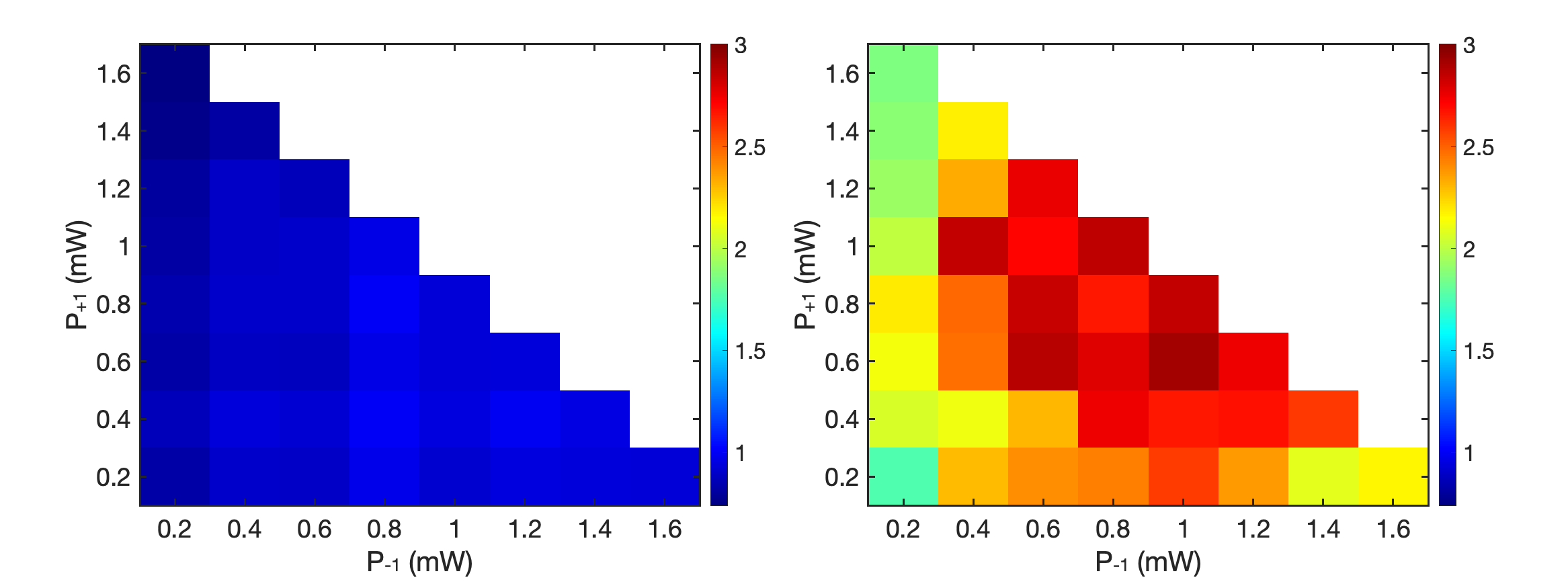}
    \caption{Plots of (a) the sum of single-frequency LIF and (b) the OC beam LIF for different combinations of $P_i$ values that satisfy a total laser beam power of 2 $\mathrm{mW}$. The color scale is identical for both plots.}
    \label{fig:PowerRatioScan2Dmap}
\end{figure}

The power ratio among the frequency components plays a critical role in determining the total scattering rate, as each transition differs in both its transition strength and the number of magnetic sublevels.
The available laser beam power is limited in practical applications such as laser slowing or MOT. 
To account for this constraint, we investigated how the power should be distributed among the frequency components to achieve efficient optical cycling. 
The total beam power $P_\text{t}$ was fixed, and various combinations satisfying $P_\text{t} = P_{-1} + P_0 + P_{+1}$ were used to measure the LIF signals (Figure \ref{fig:PowerRatioScan2Dmap}).  

The results showed that when the OC beam was applied with the optimal detunings $\delta_i$, the scattering rate was consistently higher than the sum of single-frequency LIF signals, regardless of the power ratio configuration.
The optimal power ratios for both cases were similar. 
Assigning a larger portion of the power to $P_{-1}$, which drives transitions from $F=2$ and $F=1^+$, proved to be most effective. For the remaining power, allocating it to the $F=1^-$ transition, rather than to the $F=0$ transition, yielded a higher scattering rate.
The preference for higher $P_{-1}$ can be attributed to three factors:
(1) it drives more magnetic sublevels than the other frequency components,  
(2) $\delta_{-1}$ interacts with either $F=2$ or $F=1^+$ in an off-resonant manner, requiring higher power to compensate for off-resonant scattering, and  
(3) The saturation intensity for the transition from $F=2, 1^+$ states is relatively higher than other hyperfine states due to the relatively small electric dipole matrix element.
However, as the total beam power increases and the optical cycling effect strengthens, the dominance of the $P_{-1}$ gradually diminishes.
The highest scattering rate was achieved when $P_{+1}$ and $P_{-1}$ were equal, which is attributed to the fact that the transition from $F=1^-, m_F=0$ state has the highest saturation intensity among all possible transitions.

\subsection*{Total power}
\begin{figure}[ht]
    \centering
    \includegraphics[width=0.49\linewidth]{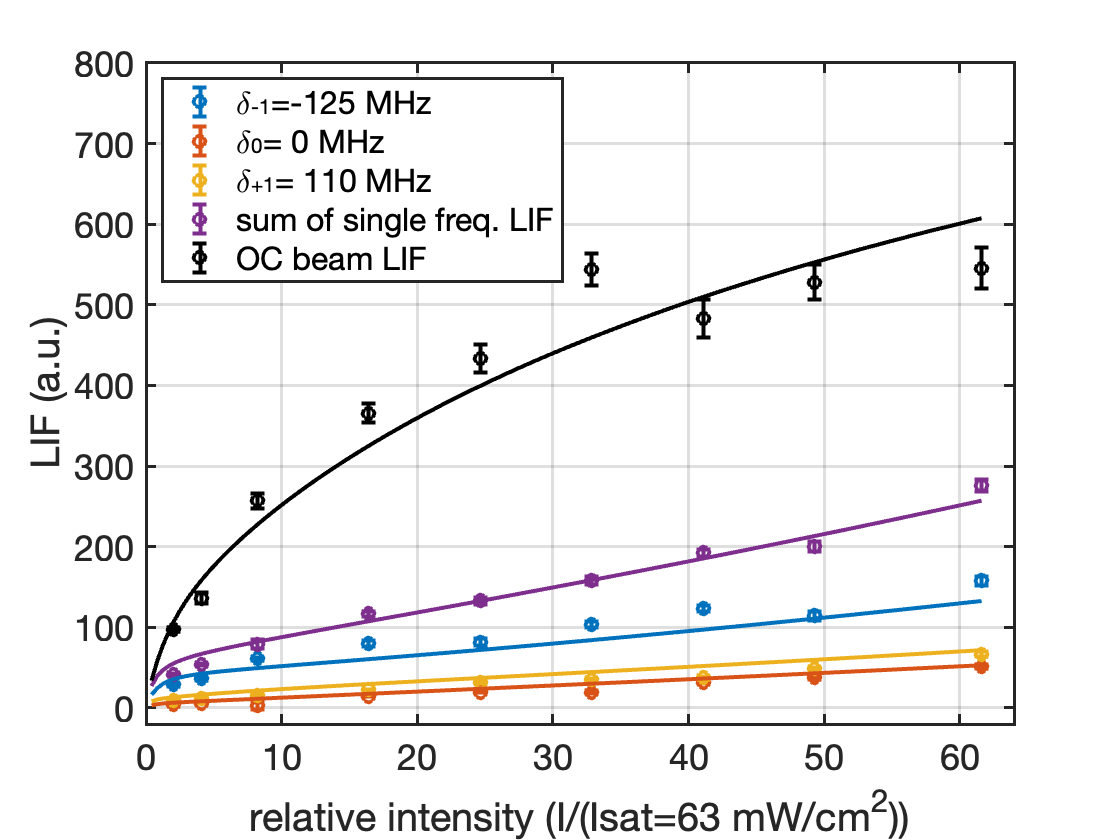}
    \caption{Total power dependence of the OC beam and single-frequency beam under the condition of $(P_{-1}:P_0:P_{+1})=(2:1:2)$. 
    The horizontal axis is given in units of the saturation intensity of the $\mathrm{X(0)-A(0)}$ band, $I_{\text{sat}}= 63\ \mathrm{mW/cm^2}$. 
    The solid line represents the result of rate equation simulations.}
    \label{fig:TotalPowerScan}
\end{figure}

Lastly, we examined the saturation behavior of the optical cycling effect as a function of the total power of the OC beam under optimal detuning and power ratio conditions.
As shown in Figure \ref{fig:TotalPowerScan}, for the single-frequency beam, the scattering rate continues to increase with beam power, as full saturation has not yet been reached.
In contrast, the LIF signal from the OC beam also increases but gradually saturates, which results from optical pumping into dark magnetic sublevels and higher vibrational states.
The solid lines in Figure \ref{fig:TotalPowerScan} represent the results of rate equation simulations, which agree well with the experimental data(See Method).
These simulations confirm that the steady-state population in the dark states increases with the total beam power.

\subsection*{Dark state mixing}

\begin{figure}[ht]
    \centering
    \includegraphics[width=0.49\linewidth]{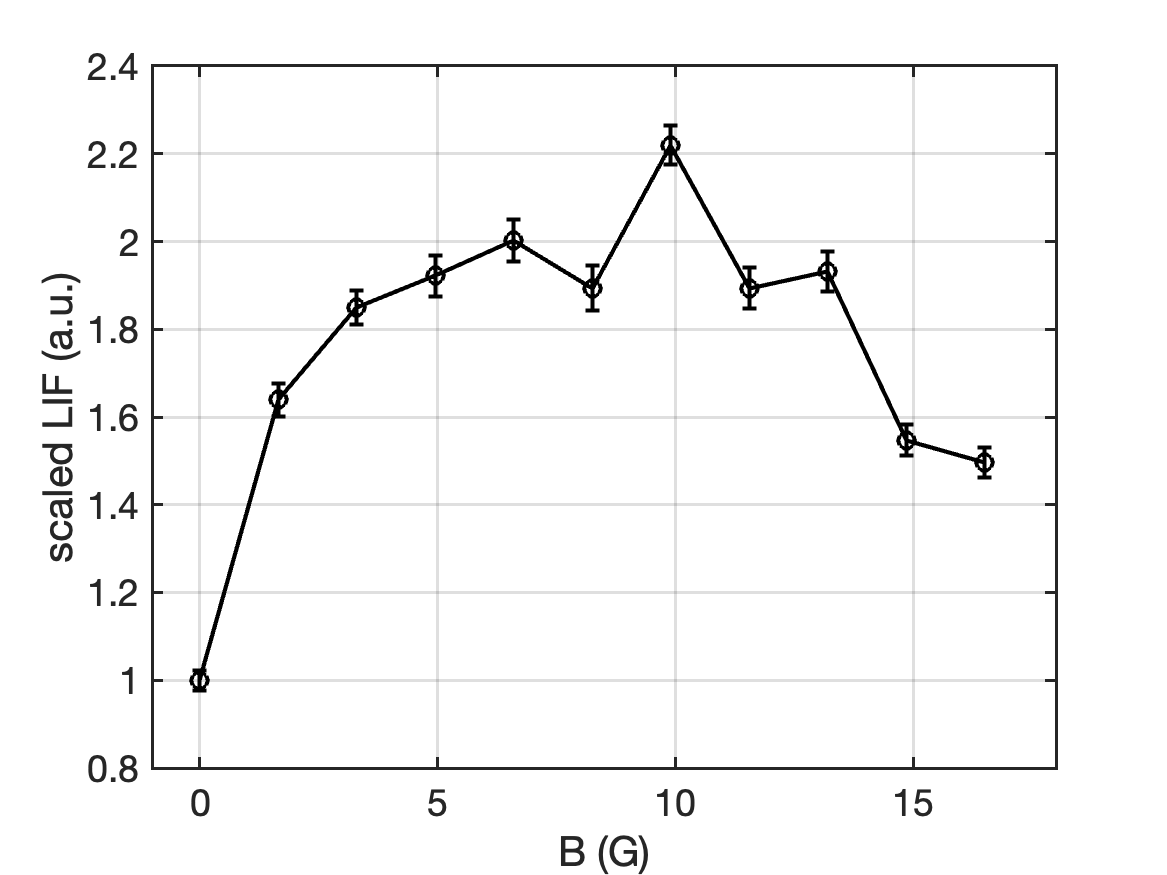}
    \caption{Plot of OC beam LIF as a function of magnetic field strength. The $y$-axis values are normalized to the OC beam LIF without a magnetic field. The detunings and power ratios of the frequency components were set to optimal conditions, and the total beam power $P_{\text{t}}$ was 2 $\mathrm{mW}$. The LIF increased with magnetic field strength from 0 to 5 $\mathrm{G}$, remained approximately constant up to 10 $\mathrm{G}$, and then began to decrease beyond that point.} 
    \label{fig:DarkStateMixing}
\end{figure}

The transitions of MgF for laser cooling are a Type-II scheme, where the number of ground states ($N=1, J=3/2, 1/2$) exceeds that of excited states ($R'=0,\ J'=1/2$), which inherently leads to the presence of dark magnetic sublevels in the ground state manifold. 
In this experiment, the laser beam was linearly polarized to drive $\pi$-transition, so the $|F=2, m_F=\pm2\rangle$ states do not have allowed transitions to the excited states.
Addressing these dark states can be achieved by modulating the laser polarization or applying a DC magnetic field at a slightly tilted angle to the laser polarization so that Larmor precession of the electron spins mixes the magnetic sublevels.

Figure \ref{fig:DarkStateMixing} shows the dependence of the OC beam LIF on the strength of the applied magnetic field $B$. 
The detuning and power ratio of the OC beam were set to their optimal values, and the magnetic field was applied at a $45^{\circ}$ angle with respect to the laser polarization. 
As the magnetic field strength increases, the LIF also increases, reaching up to 2.2 times the LIF at $B = 0$.
The optimal detuning and power ratio remained unchanged with varying magnetic field strengths.
However, at higher magnetic field strengths, the LIF begins to decrease again due to increased off-resonant scattering caused by larger Zeeman splitting and the rapid spin precession, which returns the population to dark states.

\section*{Conclusion}
We investigated the optical cycling effect on the rotationally closed $\mathrm{P_1/Q_{12}(1)}$ transition within the $\mathrm{X}(v=0) - \mathrm{A}(v'=0)$ band of the MgF molecule. 
Using AOMs, we generated three frequency components to address all relevant hyperfine transitions and independently controlled their detunings and powers. 
This approach allowed us to identify optimal parameters under which the fluorescence signal from the OC beam increased by up to a factor of three compared to the sum of individual single-frequency LIF.

Additionally, rate equation simulations showed good agreement with the experimental results, providing a consistent picture of transition dynamics and saturation behavior. 
The observed saturation at higher total beam power was attributed to population loss into dark magnetic sublevels and vibrational leakage. 
Applying a magnetic field at a non-zero angle relative to the laser polarization resulted in partial recovery of the population from dark states, leading to an additional twofold increase in fluorescence.

In summary, by fully optimizing the frequency detunings, power ratios, and incorporating magnetic field-induced dark state mixing, we achieved an overall enhancement in the scattering rate by approximately a factor of six compared to the sum of single-frequency LIF.

These results provide a basis for designing laser configurations in future MgF-based cooling and trapping experiments, including magneto-optical trapping.

\section*{Methods}
\subsection*{Experiment Setup}
\subsubsection*{Cryogenic buffer-gas beam source}

The MgF molecular beam was generated using a cryogenic buffer-gas beam source, following the structure and principle similar to most molecular laser cooling experiments~\cite{Hutzler2012_cbgb, Truppe2018_cbgb}.
Ablation of the Mg target was performed using a 532 $\mathrm{nm}$ pulsed Nd:YAG laser (Nano S 30-50, Litron Lasers) with a pulse energy of 18 $\mathrm{mJ}$ and a pulse width of <7 $\mathrm{ns}$. 
The Mg target was prepared by cutting a 99.8\% pure Mg rod into a disk shape.

The flow rate of He buffer gas influenced the main cell temperature, MgF production yield, and the forward velocity of the molecular beam. 
In this experiment, a flow rate of 3 $\mathrm{sccm}$ was optimal, as it maximized MgF production while maintaining the main cell temperature at 4 $\mathrm{K}$.

During the experiment, the sample amount gradually decreased. 
Two primary factors contributed to this. 
First, the uneven surface affected the ablation efficiency as ablation pulses removed material from the target surface. 
This issue was mitigated by randomly scanning the ablation point with a motorized mirror mount and periodically polishing the target. 
Second, the minimum flow rate of the mass flow controller (MFC) for $\mathrm{SF_6}$ gas was limited to 0.05 $\mathrm{sccm}$, which was excessively high and led to $\mathrm{SF_6}$ freezing on the inner walls of the main cell, thereby degrading the molecular beam quality.
To resolve this, the MFC was modulated using pulse width modulation, ensuring that the time-averaged flow rate was maintained at 0.01 $\mathrm{sccm}$.

The MgF molecules produced in the main cell exited as a molecular beam into the 6" CF octagonal chamber, which was connected to a beam box, where they interacted with the laser beam. 
The chamber was equipped with a turbo pump and was typically maintained at a vacuum level of $10^{-7}$ Torr with the gate valve to the beam box closed. 

\subsubsection*{Laser and optics}
The 359 $\mathrm{nm}$ laser for the main transition was generated via second harmonic generation (SHG) of a Ti:Sapphire laser (SolsTis, M Squared) operating at 718 $\mathrm{nm}$.
The SHG cavity (ECD-X, M Squared) employed an LBO crystal, achieving a maximum doubling efficiency of 30\%, yielding an output power of approximately 1 $\mathrm{W}$.

The main laser's frequency was stabilized by locking the IR frequency to a wavemeter (WS8, HighFinesse) before the SHG cavity. 
According to the factory specifications, the absolute accuracy and measurement deviation sensitivity were 10 $\mathrm{MHz}$ and 0.4 $\mathrm{MHz}$, respectively. 
The wavemeter was calibrated using a homemade ECDL locked to the D2 line of $\mathrm{^{87}Rb}$.
Calibration was performed daily, and we continuously monitored the frequency of the ECDL during each measurement to compensate for long-term drifts on the hour scale.

The power fluctuation of the main laser was stabilized using a servo system with an AOM (ISOMAT, 80 $\mathrm{MHz}$), reducing relative deviations to approximately 0.5\%.

\subsection*{Rate Equation Simulation}
The center of the forward velocity of the molecular beam generated from the CBGB is 180 $\mathrm{m/s}$. 
Given that the excited-state lifetime is $1/\Gamma=7.2\ \mathrm{ns}$, 
The rate equation can sufficiently describe the interaction process while the molecules traverse a laser beam with a waist radius $w_0=0.5$ $\mathrm{mm}$~\cite{Wall2008_CaF_lifetime_our_simulation}. 
The rate equations used in the simulation are given by:

\begin{align}
\frac{dN_g}{dt}&=\sum_{e,i} R_{eg,i}(N_e-N_g)+\Gamma f_{00} \sum_eB_{eg}N_e\\
\frac{dN_e}{dt}&=\sum_{g,i} R_{eg,i}(N_g-N_e)-\Gamma N_e\\
\frac{dN_{v'}}{dt}&=\Gamma (1-f_{00}) \sum_e N_e
\end{align}
Where $N_g$, $N_e$, and $N_{v'}$ represent the populations of the ground-state magnetic sublevels, excited-state magnetic sublevels, and vibrationally excited states ($v>0$), respectively. 
Here, $f_{00}$ is the branching fraction of the $\mathrm{X}(0)-\mathrm{A}(0)$ band, and $B_{eg}$ denotes the branching ratio determined by Clebsch-Gordan coefficients.

The transition rate $R_{eg,i}$ for the $i$-th frequency component of the OC beam, considering its detuning $\delta_i$ and power $P_i$, is expressed as:

\begin{align}
&R_{eg,i} = \frac{\Gamma/2}{1+4(\delta_{i}-\Delta_{eg})^2/\Gamma^2}\frac{I_i(t)}{I_{eg}}\\
&I_{eg} = \frac{\pi hc}{\lambda^3}\frac{\sum_{g'}m_{eg'}^2}{m_{eg}^2}
\end{align}

where $\Delta_{eg}$ represents the hyperfine energy splitting, $I_{eg}$ is the saturation intensity, and $m_{eg}$ is the electric dipole moment matrix element.

Since the laser beam has a Gaussian intensity profile, the time-dependent intensity $I(t)$ experienced by the molecule traveling through the laser beam must be accounted for at each time step. The time range for solving the rate equations is set to scan from $-2w_0$ to $2w_0$ relative to the intensity peak, resulting in a total scan time of $4w_0/v_f$.

The Doppler broadening due to the molecular beam's transverse velocity distribution could be ignored, as power broadening was the dominant effect in our experiment.

\section*{Data availability}
All data included in this study are available upon request by contacting the corresponding author, Eunmi Chae (echae@korea.ac.kr).

\bibliography{references}

\section*{Acknowledgements}
Authors appreciate Donghyun Cho for his expert advice and for providing essential experimental equipment.
We are especially grateful to Dongkyu Lim, whose preliminary calculations played a key role in our simulation framework.
We also thank Hyunjun Jang, Hyeongtae Kim, and Zhirui Cheng for their assistance with the experimental setup.
The authors acknowledge support from the National Research Foundation of Korea under grant numbers RS-2022-NR119745,  RS-2024-00439981, RS-2024-00431938,  and RS-2023-NR068116.

\section*{Author contributions statement}
E.C. supervised the project and set the direction of the experiment. 
K.K., S.R., Y.C., and Y.L. designed and built the experimental setup, carried out the measurements, and analyzed the data.
K.K., Y.C., and Y.L. performed the simulation calculations. 
K.K. wrote the original draft. 
All authors reviewed and edited the manuscript.

\section*{Additional information}
\subsection*{Competing interests}
The authors declare no competing interests.

\section*{Legends}
\begin{itemize}
    \item \textbf{Figure 1.} (a) Quasi-closed cycling transitions for the rovibrational branches of MgF. Solid lines represent the transitions and their wavelengths, while dashed lines indicate dominant vibrational decay channels with their branching fractions. 
    (b) Fine and hyperfine energy levels of the $\mathrm{X}(v=0,\ N=1^-)-\mathrm{A}(v'=0,\ J'=1/2^+)$ band. The black arrows indicate the rotationally closed transitions, $\mathrm{P_1/Q_{12}(1)}$, while the blue arrows represent the frequencies of the excitation lasers for $\mathrm{P_1/Q_{12}(1)}$ transition. The detuning of each laser is referenced to 834294485 $\mathrm{MHz}$, which is the frequency of the transition from $F=0$ state.
    
    \item \textbf{Figure 2.} (a) Schematic diagram of the experimental setup. Two AOMs generate the frequency components of the OC beam, which is delivered to the chamber via an optical fiber. The molecular beam(yellow region), produced by the CBGB(Cryogenic buffer-gas beam) source, interacts perpendicularly with the laser beam approximately 34 ${\mathrm{cm}}$ downstream. The PMT records the LIF signal for 20 $\mathrm{ms}$ after ablation.
    (b) Spectra for the OC beam (solid black line) and single-frequency components (dashed line). LIF signals were obtained by scanning the main laser frequency while keeping $(\delta_i,\ P_i)$ of each frequency component fixed. The purple solid line represents the sum of single-beam spectra at each frequency. Near resonance, where all three frequency components address the hyperfine transitions, the optical cycling effect becomes evident.

    \item \textbf{Figure 3.} (a, b) The dependence of LIF from the OC beam as a function of (a)$\delta_{-1}$  and (b)$\delta_{+1}$, while keeping the other components fixed at (a) ($\delta_{0}=0$, $\delta_{+1}=110$) MHz and (b) ($\delta_{-1}=-125$, $\delta_{0}=0$) MHz, respectively. 
    Solid lines indicate simulation results, and dashed lines mark transition frequencies for $F=2$, $F=1^+$, and $F=1^-$. (c) Comparison of LIF time traces between the OC beam at the optimal detuning and the single-frequency beam. All experiments were conducted with a total laser beam power of 2 $\mathrm{mW}$, distributed in the ratio of $(P_{-1}:P_0:P_{+1}) = (5:2:3)$.

    \item \textbf{Figure 4.} Plots of (a) the sum of single-frequency LIF and (b) the OC beam LIF for different combinations of $P_i$ values that satisfy a total laser beam power of 2 $\mathrm{mW}$. The color scale is identical for both plots.

    \item \textbf{Figure 5.} Total power dependence of the OC beam and single-frequency beam under the condition of $(P_{-1}:P_0:P_{+1})=(2:1:2)$. 
    The x-axis is expressed in units of the saturation intensity of the $\mathrm{X(0)-A(0)}$ band, which is 63 $\mathrm{mW/cm^2}$. 
    The solid line represents the result of rate equation simulations.
    
    \item \textbf{Figure 6.} Plot of OC beam LIF as a function of magnetic field strength. The $y$-axis values are normalized to the OC beam LIF without a magnetic field. The detunings and power ratios of the frequency components were set to optimal conditions, and the total beam power $P_{\text{t}}$ was 2 $\mathrm{mW}$. The LIF increased with magnetic field strength from 0 to 5 $\mathrm{G}$, remained approximately constant up to 10 $\mathrm{G}$, and then began to decrease beyond that point.
\end{itemize}

\end{document}